\documentclass{ringb99}
\usepackage{graphics}
\def\h5{\h_{50}}
\def\beq{\begin{equation}}
\def\eeq{\end{equation}}

\def\hide#1{}
\def\pic#1{#1}
\begin{document}
\titlerunning{74 MHz-VLA-Observation of Coma Berenices with Subarcminute
  Resolution}
\title{74 MHz-VLA-Observation of Coma Berenices with Subarcminute
  Resolution -- Observation, Data Reduction, and Preliminary Results}
\author{T.A. En{\ss}lin\inst{1,2} \and P.P. Kronberg\inst{1} \and R.A.
  Perley\inst{3} \and N.E. Kassim\inst{4}}

\institute{Department of Physics, University of Toronto, 60 St. George
  Street, Toronto, ON, M5S1A7, Canada \and Max--Planck--Institut f\"ur
  Radioastronomie, Auf dem H\"ugel 69, 53121 Bonn, Germany \and
  National Radio Astronomy Observatory, PO Box 0, Socorro, NM 87801,
  USA \and Naval Research Laboratory, Code 7213, Washington, DC, USA }

\maketitle

\begin{abstract}
  We report on an ongoing project to map a $15^\circ\times15^\circ$
  region in Coma Berenices with the new 74 MHz receiver system of the
  VLA. The field is centered on the Coma cluster of galaxies. Coma
  Berenices has been  observed in all VLA configurations, which allowed us
  to resolve several hundred point sources with 35 arcsecond
  resolution.  We have successfully detected the diffuse emission from
  the Coma cluster radio halo, radio relic, and radio bridge.
  Difficulties and strategies of the observation and data reduction
  are discussed. Preliminary results are given.
\end{abstract}

\section{Scientific Goals}
The Coma cluster of galaxies contains one of the largest known radio
halos, a prominent radio relic and a so called `radio bridge' between
them. Although it is one of the best studied clusters of galaxies, the
processes leading to these diffuse radio emissions are not completely
understood. The different proposed explanations are well covered by
articles of several authors within the proceedings of this conference.

Since the diffuse radio emission has a steep spectral index ($\alpha >
1$) and seems to be more extended at lower frequencies, a 74 MHz-VLA
observation seemed to be a promising project. The goals are:

\begin{itemize}
\item[$\circ$] to get information about the relativistic, low energy
  electron and magnetic field distribution in the Coma cluster
\item[$\circ$] to search for extended emission outside of Coma,
  possibly connected with the {\it `Great Wall'}, a galaxy filament
  which contains the Coma cluster, and which extends over 100 Mpc.
\item[$\circ$] to obtain fluxes and spectral indices  of sources located in the field.
\item[$\circ$] to search for ultra-steep spectrum sources, which would
  only appear at lowest frequencies.
\item[$\circ$] to test and demonstrate the abilities of the new 74 MHz
  receiver system installed on all 27 VLA antennas. An earlier version
  installed on eight antennas had been successfully tested only with
  very bright sources (Kassim et al.  1993). But it was not yet clear
  how sensitive the system would be, or in
  our case, whether the faint diffuse emission of a cluster radio halo 
could be detected.  
\item[$\circ$] to develop an optimal observing and data reduction
  strategy for 74 MHz.  \end{itemize} Observations in A-, B-, C-, and
D-configuration of the VLA have been successfully undertaken. The data
analysis is still in progress, but it is clear at the present stage
that several of the goals will be accomplished.

\section{Very Low Frequency Radio Observation}
Simultaneous observations at 74 MHz and at 328 MHz were carried out.
The latter frequency gives additional spectral information. It was
also chosen to enable calibration by the phase-transfer method in case
phase calibration at 74 MHz proved difficult. The details of this
method are described in Kassim et al.  (1993), who also give a
general introduction to observations with the 74 MHz VLA receiver
system. It turned out that the phase-transfer method was not
necessary, since a sufficient number of strong sources located in the
observing field allow for easy self-calibration (Pearson \& Readhead
1984) of the 74 MHz data. But this need not to be always true: In the
case of recent VLA galactic center observations the phase-transfer
method was required to calibrate the 74 MHz data.

We observed in the VLA's spectral line mode (128 channels for the 74
MHz observation and 64 channels for the 328 MHz observation) for two
reasons:
\begin{itemize}
\item[$\circ$] radio frequency interference, externally generated by
  strong natural and artificial emitters and internally generated by
  the receiver electronics, is strong at low
  frequencies at the VLA.  However it is generally very narrow-band and so the
  affected channels can be flagged.
\item[$\circ$] the total bandwidth of 1.6 MHz leads to excessive bandwidth
  smearing over our large field of view. Imaging the data in spectral
  line mode allows to compensate for this.
\end{itemize}

\section{Very Low Frequency Data Reduction}
A tutorial for new users of the 74 and 330 MHz systems at the VLA has
been written up by Namir Kassim \& Rick Perley. It is available from\\[0.5em]
{\tt http://rsd-www.nrl.navy.mil/7213/lazio/tutorial}\\[0.5em]
and in the calibration sections of the NRAO VLA web pages through:\\[0.5em]
{\tt http://www.nrao.edu/}\\[0.5em]

There are three main complications with low frequency radio
observations at the VLA, which are less problematic at higher frequencies: the
large primary beam, the impact of the ionosphere, and the
strong radio frequency interference. New features of present (AIPS,
Miriad), and upcoming software packages (AIPS++) help to deal with
these problems.

\subsection{Radio Frequency Interference}
\begin{figure}
  \pic{\resizebox{\hsize}{!}{\includegraphics{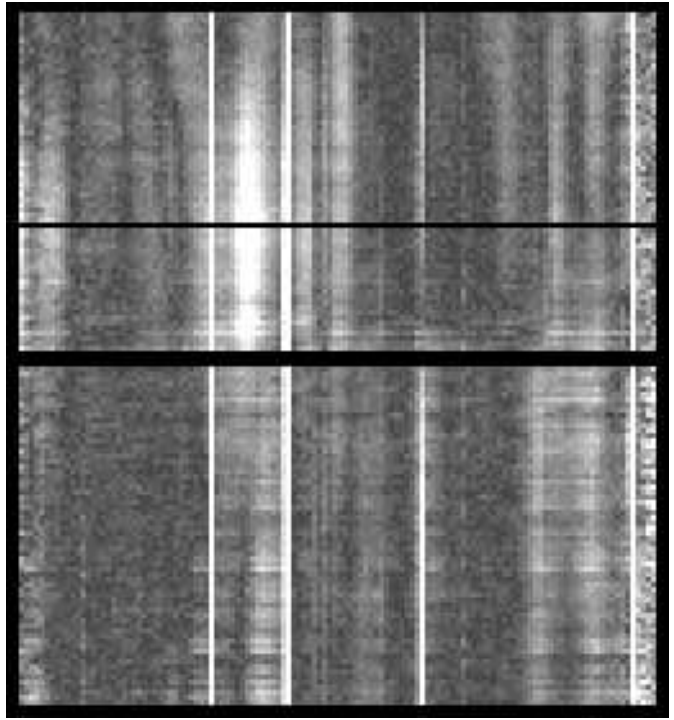}}}
  \caption[]{\label{fig:rfi}
    Plot of the cross-correlation amplitude of the antennas at the
    western and northern end of the VLA array in D configuration. The
    x-axis is the spectral line channel number and the y-axis the
    observing time. Each of the two blocks corresponds to 24 min
    observation. The worst RFI in this plot (white lines) is 100 times
    stronger than the astronomical signal (dark grey regions). Note
    that although this example was chosen for its very strong RFI, it
    contains regions of usable data.}
\end{figure}
\begin{figure*}
  \pic{\resizebox{\hsize}{!}{\includegraphics{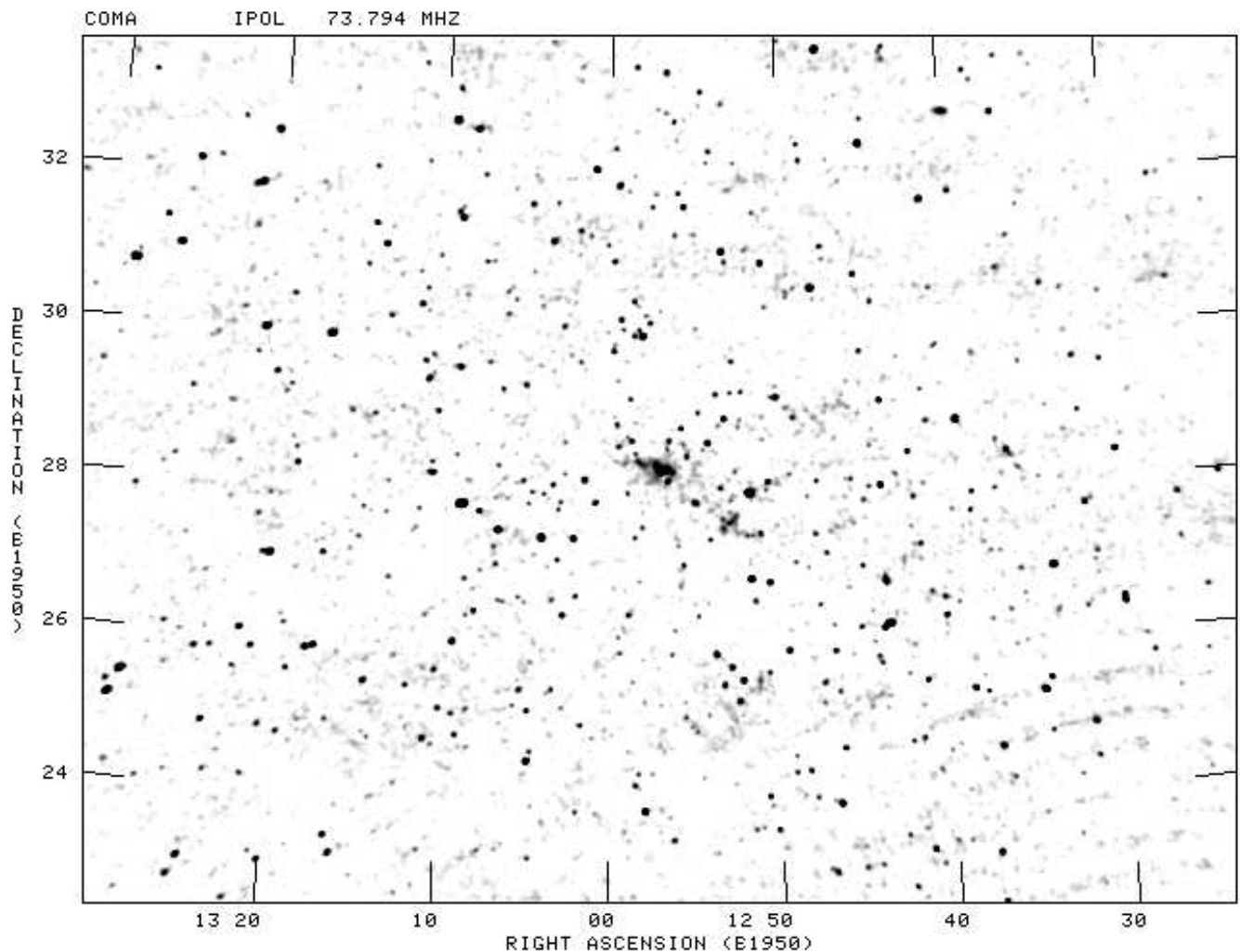}}}
  \caption[]{\label{fig:field}
    $16^\circ\times12^\circ$ field of view, which covers a large
    fraction of the primary beam. The image is smoothed to a beam size
    of $200\arcsec\times200\arcsec$ in order to make the diffuse
    emission of the Coma cluster of galaxies visible. The field
    contains several hundred point sources. Note the diffuse rings,
    which are most prominent in the lower right corner. These are
    sidelobes from Virgo A, which is located $17^\circ$ from the
    center of the field. Strategies to remove these sidelobes are
    discussed in the text.}
\end{figure*}

Radio frequency interference (RFI) can be external or internal, but is
usually of very narrow bandwidth, so that it is easy to detect in a
spectral line data base. The most recent version of AIPS has a new
task (FLGIT) which greatly facilitates removal of RFI. In general, RFI
is worst for the shortest baselines. An important test for the VLA 74
MHz system is to see if the data taken with compact array
configurations can be used, despite the stronger RFI.  The B- and C-
configuration data from Coma have been successfully reduced. The data
for the most compact D-configuration is not yet reduced, so final
conclusions cannot be drawn.  But the present results look very
promising. An example of strongly RFI-affected data can found in Fig.
\ref{fig:rfi}.

\subsection{Ionospheric Effects}

Ionospheric turbulence on size scales of several km enters as
time-variable antenna based phase variations, and decorrelates the
signals on long baselines for the 74 MHz VLA. This simple fact has
prevented high angular resolution and high sensitivity imaging on
interferometers working at low frequencies ($< 100$ MHz) since radio
astronomy was developed, and as a result this region remains one of
the poorest explored regions of the electro-magnetic spectrum. Hence
the recent demonstration that phase self-calibration can correct this
is important (Kassim et al. 1993). Self-calibration works robustly for
74 MHz VLA observations since typical fields of view contain numerous
strong sources. This requires that the data have a sufficient time
resolution. For A-configuration data the time integration interval
should not exceed 10~s, and for the more compact configurations
correspondingly longer.  The same time resolution is required in any
case to prevent the peripheral sources becoming smeared out due earth
rotation.

Unfortunately, this ionospheric effect is different for the different
parts of the image, due to different lines of sight through the
ionosphere. Thus some miscalibration for some regions of the primary
beam (typically peripheral regions) cannot be avoided. This problem
might disappear when direction dependent self-calibration is possible
in up-coming reduction software.

Observing at low elevations should be avoided, since the line of sight
through the ionosphere is longest. Periods in which the ionosphere is turbulent 
also might best be removed completely from the database.

Finally, ionospheric refraction has an especially serious effect on
the astrometry of 74 MHz VLA images, shifting the field of view though
not distorting the brightness distribution (Erickson 1984). Shifts at
74~MHz are typically a few arcminutes and vary on time scales of tens
of minutes.  Self-calibration has no trouble `freezing out' this
refraction but leaves the image with an uncertain absolute position.
Fortunately the grid of many tens to hundreds of known sources which
normally appear on 74 MHz VLA images allows us to re-register the
absolute position quite accurately relative to our angular resolution.

\begin{figure*}
  \pic{\resizebox{\hsize}{!}{\includegraphics{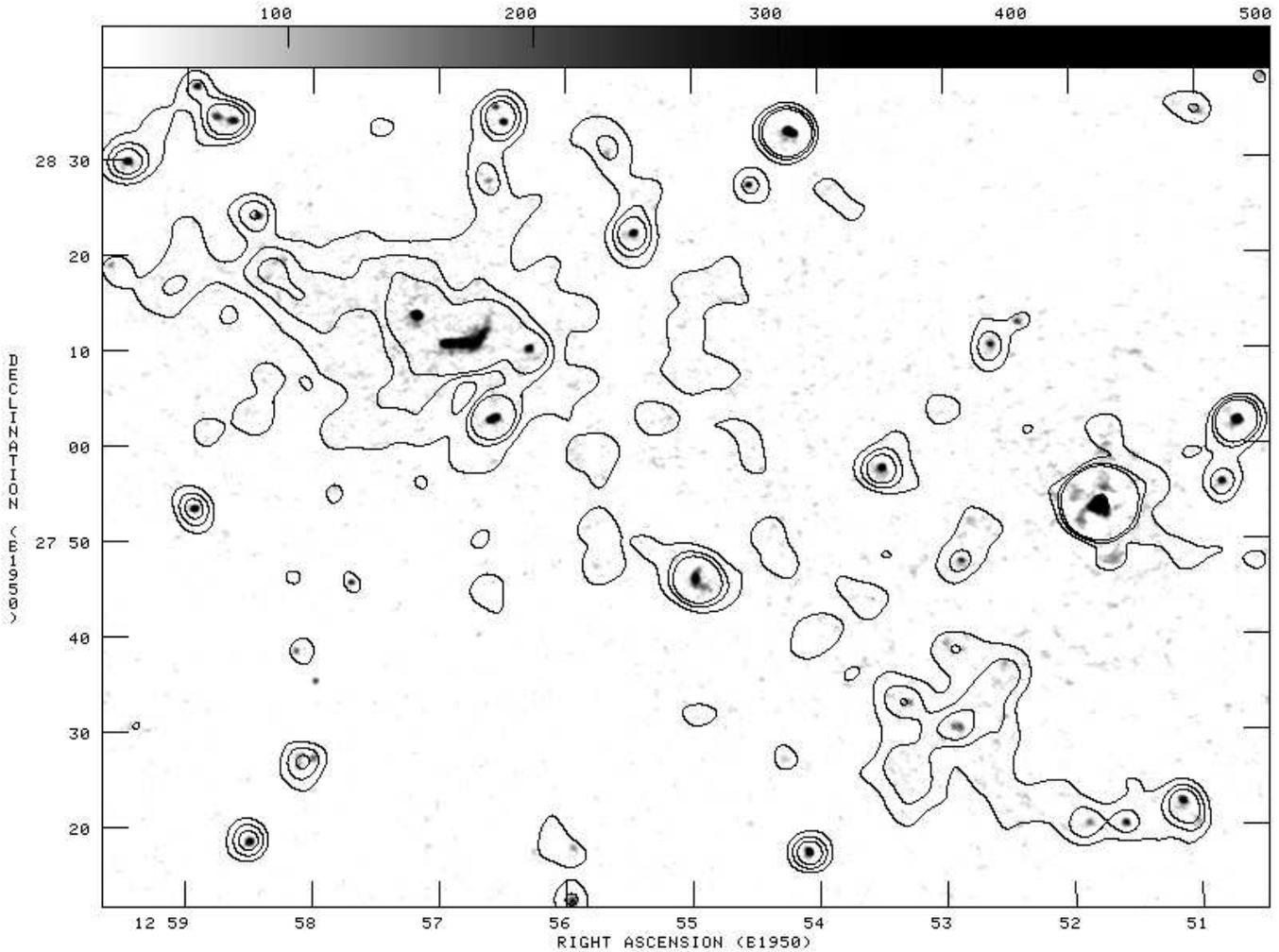}}}
  \caption[]{\label{fig:center}
    Coma cluster of galaxies with the full resolution of $35\arcsec$
    of the A-, B-, plus C-configuration data in grey scale.  The contours
    are 8, 16, 24 mJy/arcmin$^2$ levels of an image smoothed to a
    $200\arcsec\times200\arcsec$ beam. The radio halo Coma-C and the
    radio relic 1253+275 of Coma are prominent, the radio bridge
    connecting them is also detected.}
\end{figure*}
\begin{figure*}
  \pic{\resizebox{\hsize}{!}{\includegraphics{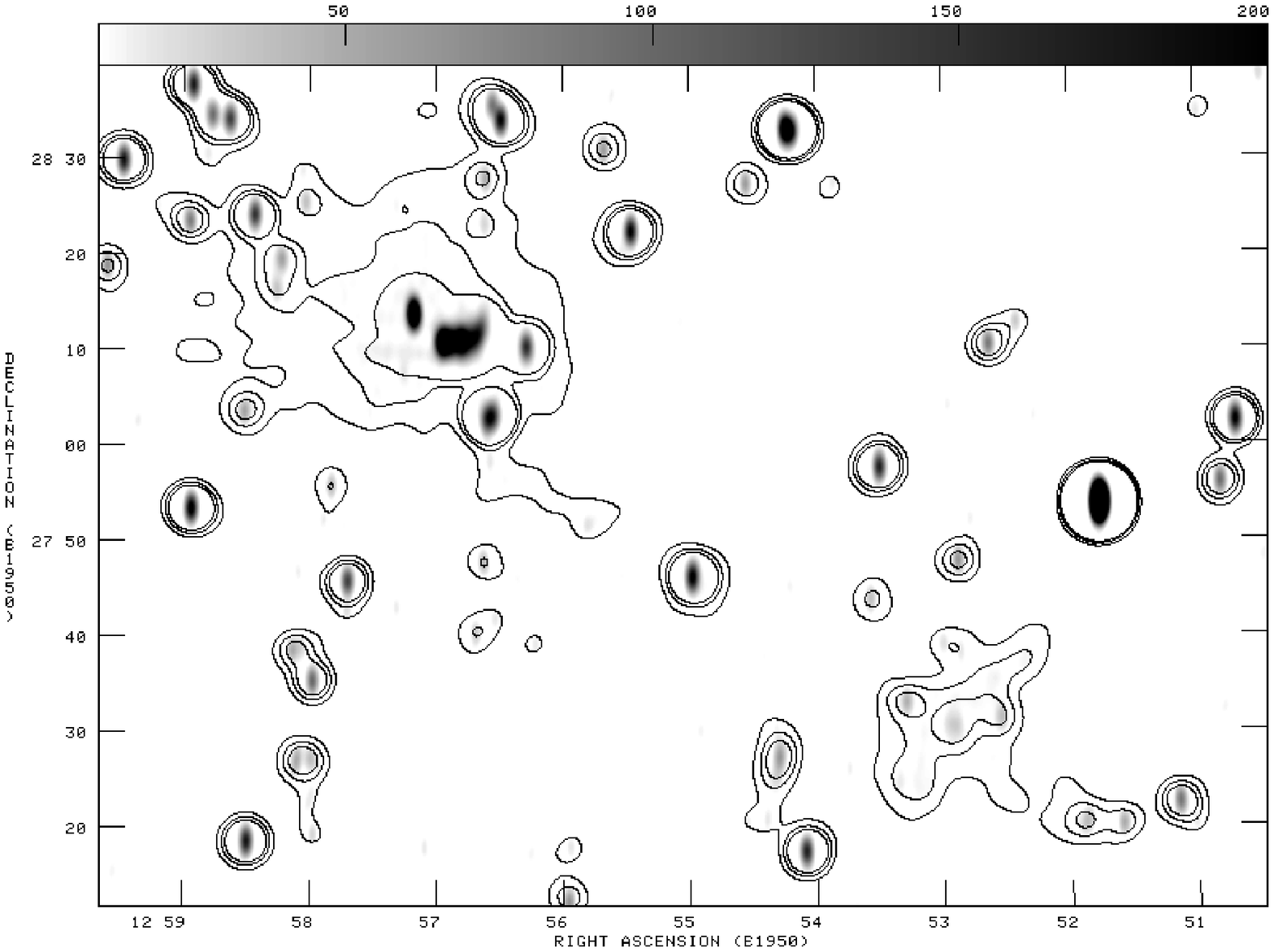}}}
  \caption[]{\label{fig:centerW}
    Coma cluster of galaxies observed with the Westerbork telescope at
    326 MHz (Kim et al. 1989). The grey scale image is the full
    resolution of $127\arcsec\times 55\arcsec$, and the contours image
    is the smoothed emission with a $200\arcsec\times 200\arcsec$
    beam. The smoothing and the contour levels of 4, 8, 12
    Jy/arcmin$^2$ are chosen to allow a morphological comparison with
    the 74 MHz diffuse emission in Fig.  \ref{fig:center}. Note that the
    elliptic appearance of the point sources is due to the elliptical
    synthesized beam of the Westerbork array.}
\end{figure*}
\subsection{The Large Primary Beam}
The large primary beam is beneficial, since it allows one to map a
large fraction of the sky, but it causes several problems:
\begin{itemize}
\item[$\circ$] The large primary beam requires mapping of a large field,
  in order to reduce the sidelobes of strong sources. This requires a
  high computational effort.
\item[$\circ$] The usual (fast) two dimensional fourier transform
  between UV- and image-plane approximates the sky by a (tangential)
  plane. For non-coplanar baselines, as with the VLA, this would lead
  to serious errors for a field of view as large as ours. The solution
  is either the use of a real three-dimensional code, or to split
  the field into small facets, which are small enough to let the planar
  approximation be valid (Cornwell \& Perley 1992). We tried both
  approaches, but since the second method gives good results with a
  much smaller computational effort, we concentrated on the facet
  method. We used up to 64 facets, the limit of the present AIPS
  version\footnote{The 15OCT99 version of AIPS  allows 512
    facets.}, which are not enought to cover the whole field for
  A-configuration resolution.

\item[$\circ$] The units of the UV-plane are baseline length divided
  by wavelength. These change over the bandwidth, so that
  averaging the spectral line data over the whole band prior to
imaging leads to radial smearing of
  sources in image plane. The data therefore must be
  kept in spectral line mode, in order to suppress this
  bandwidth-smearing. However, the number of channels can be reduced by
  averaging, to keep the computational effort tolerable.
\item[$\circ$] The array is sensitive to sources within a significant
  fraction of the sky, much larger than the primary beam. In our case
  Virgo A is located in the sidelobe of the VLA beam. Since the
  sidelobe pattern is non-axisymmetric, the gains of the telescopes
  for this source change with time. This makes a mapping and cleaning
  of the source difficult, with the result that it has large sidelobes
  all over the image (Fig.  \ref{fig:field}). We are currently
  pursuing two strategies in order to remove Virgo A from our dataset:
  \begin{itemize}
  \item The information about the time dependent antenna gains is
    contained in the data itself. It should be possible to do a full
    (phase \& gain) calibration on a good model of Virgo A, subtract
    it and go back to the original calibration. This should account
    for the time varying gains, since they will be estimated during
    the amplitude calibration.
  \item Sault \& Noordam (1995) have developed a method to remove
    continuum interference from sidelobe responses from a spectral
    line dataset.  This method might be applicable in our case.
  \end{itemize}
\end{itemize}

\section{Preliminary Results}
Self calibration of the data over the large field of view of
$15^\circ\times 15^\circ$ was successful, as can be seen from the
image in Fig.  \ref{fig:field}. A large number of point sources are
visible, on the order of several hundred.  Diffuse emission from
Coma's radio halo, radio relic and radio bridge was also detected.
Combining A-, B-, and C-configuration data leads to a high resolution
map, which we show in superposition with a smoothed contour map from
the same data in Fig. \ref{fig:center}.  The theoretical resolution of
$25\arcsec$ is has not been achieved yet, but the current resolution
of $35\arcsec$ demonstrates the ability of the new VLA receiver
system, and the self-calibration procedures to compensate for the
ionospheric fluctuations. This map is preliminary and we expect that a
further improvement in resolution will be achieved.

The level of noise in the high resolution image is 20 mJy/beam,
reached with 12 h integration time in each of the three
configurations.  For comparison, the brightness of the radio halo is
40 mJy/beam.

For comparison, we show a map of the 326 MHz emission measured with
the Westerbork telescope by Kim et al. (1989). Fig.  \ref{fig:centerW}
was prepared in a similar way to Fig.  \ref{fig:center}. Several of
the irregular features of the radio halo of Coma presented in the
Westerbork map are confirmed by our 74 MHz observation. Note that
diffuse emission on larger scales might still be missing on our 74 MHz
map, since the UV-coverage is not complete.  With C-configuration data
included, we are sensitive to diffuse emission on scales less than
$\sim 1^\circ$.  We expect that our D-configuration data will allow us
to test for diffuse emission on scales up to $3^\circ$. This is very
interesting, since it is closer to the diameter of the `Great Wall', a
galaxy filament in which Coma is embedded. If there is diffuse low
frequency radio emission connected to it, it might show up in this
observation.

\section{Conclusion and Outlook}

We have successfully mapped Coma Berenices at 74 Mhz with 3 VLA
configurations, and have data from the most compact one. The
preliminary results are very encouraging, but more effort needs to be
spent in understanding the artifacts, sidelobes of strong sources, and
determining the optimal data reduction strategy.  As the required
software tools come on line, this low frequency system is now
developing into a standard observing band at the VLA.  Future
increases in computational speed, especially by the use of parallel
codes on multiprocessor computers, hopefully will allow a much faster
reduction.
 
Scientific results relevant to the diffuse emission from the Coma
cluster of galaxies can be expected soon, since the quality of our
data is surprisingly good. Also, the fluxes of the several hundred
point sources will be measured, giving us a survey of sources above
100 mJy in the Coma Berenices region with a single telescope pointing.

We hope that this study encourages more research in low frequency
radio astronomy.

\begin{acknowledgements}
  We gratefully acknowledge Peter L. Biermann, Luigina Feretti,
  Gabrielle Giovannini, and Bob Hanisch for their work on the
  observation proposal. We are thankful to Tim Cornwell, Ron Ekers,
  Bill Erickson, Frazer Owen, Bob Sault and Greg Taylor for help and
  suggestions.  This work took great advantage of recently developed
  software by Eric Greisen, whom we thank for this and his advice. We
  especially thank Jim Condon to make it possible to obtain VLA
  D-configuration data.  TAE acknowledges support from the {\it
    National Science and Engeneering Research Council of Canada}
  (NSERC), the {\it Max-Planck-Institut f\"ur Radioastronomie}
  (MPIfR-Bonn) and the {\it Studienstiftung des deutschen Volkes}.  He
  further acknowledges the warm hospitality of the {\it Array
    Operation Center} (AOC-Socorro) of the VLA, where most of the data
  reduction was done up to now. Basic research in radio astronomy at
  the {\it Naval Research Laboratory} is supported by the {\it Office
    of Naval Research}.
\end{acknowledgements}

\hide{
  \begin{table}
    \caption{Example of a table.}
    \label{KapSou}
    \[
    \begin{array}{p{0.5\linewidth}r}
      \hline
      \noalign{\smallskip}
      Type of Contribution    &  {\rm No. Pages} \\
      \noalign{\smallskip}
      \hline
      \noalign{\smallskip}
      talks                  &  6 \\
      posters                &  4 \\ 
      \noalign{\smallskip}
      \hline
    \end{array}
    \]
    \begin{list}{}{}
    \item[$\circ$][$^{\rm a}$] This is a footnote
    \end{list}
  \end{table} 
  } 

\hide{
  \begin{figure}
    \pic{\resizebox{\hsize}{!}{\includegraphics{test_figure.ps}}}
    \caption[]{Example of an included figure ( .ps file).}
  \end{figure}
  }
\end{document}